\documentclass[preprint, sort&compress, 12pt]{elsarticle}

\usepackage{amssymb}

\newcommand{\average}[1]{\ensuremath{\langle#1\rangle} }

\journal{Physica A}

\begin{document}

\begin{frontmatter}



\title{Effects of updating rules on the coevolving prisoner's dilemma}


\author{Hirofumi Takesue}

\address{Graduate Schools for Law and Politics, The University of Tokyo\\
7-3-1, Hongo, Bunkyo, Tokyo, 1130033, Japan\\}
\ead{hir.takesue@gmail.com}

\begin{abstract}
We studied the effect of three strategy updating rules in coevolving prisoner's dilemma games where agents (nodes) can switch both the strategy and social partners. Under two node-based strategy updating rules, strategy updating occurs between a randomly chosen focal node and its randomly selected neighbour. The focal agent becomes the strategy recipient and may imitate the strategy of the neighbour according to the payoff difference, i.e. voter-model-like dynamics (VMLD), or becomes a strategy donor and thus may be imitated by the neighbour, i.e. invasion-process-like dynamics (IPLD). For edge-based updating rules, one edge is selected, and the roles of the two connected nodes (donor or recipient) are randomly decided, i.e. edge-based dynamics (EBD). A computer simulation shows that partner switching supports the evolution of cooperation under VMLD, which has been utilised in many studies on spatial evolutionary games, whereas cooperators often vanish under IPLD. The EBD results lie between these two processes. This difference is prominent among nodes with large degrees. In addition, partner switching induces a non-monotonic relationship between the fraction of cooperators and intensity of selection under VMLD and EBD, and a weak or strong selection supports cooperation. In contrast, only a strong selection supports cooperators under IPLD. Similar differences in the enhancement of cooperation are observed when games are played on static heterogeneous networks. Our results imply that the direction of imitation is quite important for understanding the evolutionary process of cooperation.
\end{abstract}


\end{frontmatter}


\section{Introduction}
The evolution of cooperation is an actively studied subject in physical and biological science \cite{Nowak2006, Szabo2007a, Perc2017}. In social interactions, cooperators must pay a cost for the benefit of others. Despite the benefit of mutual cooperation, natural selection appears to hinder the evolution of cooperation because non-cooperative individuals can receive the benefit of cooperation without bearing the cost of a cooperative act. The prisoner's dilemma is a widely adopted framework that represents this social dilemma. In the prisoner's dilemma game, two players (agents) simultaneously choose whether to cooperate ($C$) or defect ($D$). They will receive $R$ if both choose cooperation and $P$ if both choose defection. If one player cooperates and the other one defects, the cooperator receives $S$ and the defector receives $T$. Because the order of the payoff is $T > R > P > S$, players should choose $D$ regardless of the partner's choice if they wish to maximise their own payoff. This temptation leads to mutual defection, although the realised payoff ($P$) is smaller than the result of mutual cooperation ($R$). However, this prediction contradicts the widely observed cooperation in actual human society.

Many models have been proposed to study the evolutionary origin of cooperation, including the effect of the network (spatial) structure. In their pioneering work, Nowak and May \cite{Nowak1992} showed that cooperation proliferates if the players are located on a two-dimensional lattice. Subsequent studies introduced complex networks that incorporate the properties of actual networks, such as high clustering and degree heterogeneity, and defined the effects of them on the evolution of cooperation. Notably, a scale-free network gives a unified explanation on the emergence of cooperation in the prisoner's dilemma as well as other games \cite{Santos2005, Santos2006b, Santos2006a}. These studies highlight the importance of the heterogeneity in degree (i.e. the number of neighbours of each node). Following works examined the robustness of this phenomenon under wider conditions \cite{Masuda2007, Wu2007, Szolnoki2008, Yang2012}. In addition, the effect of the network structure on the evolution of cooperation was investigated in other networks, including random regular graphs \cite{Vukov2006}, small world networks \cite{Abramson2001, Kim2002, Masuda2003} and actual social networks \cite{Holme2003, Fu2007b, Wu2015a}. Furthermore, the role of networks in resolving social dilemma was investigated in combination with other mechanisms, including voluntary participation \cite{Szabo2002, Wu2005}, heterogeneous teaching activity \cite{Szolnoki2007, Szolnoki2008c, Wu2015b}, time scale for strategy updating \cite{Wu2009d, Rong2010, Rong2013}, payoff aspiration \cite{Chen2008, Liu2011, Xu2017a}, conformity \cite{Cui2013, Szolnoki2015, Javarone2016b} and punishment \cite{Helbing2010d, Szolnoki2011a, Chen2015, Yang2015b}.

In addition to analysing the effect of static networks, recent literature analysed a coevolutionary game where both the network structure and the players' strategy evolve. In coevolutionary games, agents can sever the relationship with a current neighbour and construct a new link with other agents \cite{Zimmermann2004, Pacheco2006a, Santos2006, Fu2007, Fu2008, VanSegbroeck2008, Fu2009, Meloni2009, Perc2010, Yang2013a, Cardillo2014, Cong2014, Xu2014, Chen2016, Pinheiro2016, Wang2016f, Li2017b}. In many of these models, the criterion of the continuation of the relationship depends on the agents' strategy or payoff from the game. These studies showed that the possibility of partner switching (link adaptation) greatly enhances the evolution of cooperation compared to static graphs. The effect of coevolution was also studied with other games, including the snowdrift game \cite{Graser2009, Zhang2014}, stag hunt game \cite{Zhang2016a} and ultimatum game \cite{Deng2011, Gao2011, Takesue2017a}.

In many of these studies, it is assumed during strategy evolution one randomly chosen \textit{focal} agent (node) decides whether to imitate the strategy of a randomly chosen neighbour by comparing their payoff from games \cite{Wu2009b}. This means that the role of a focal agent is fixed to a strategy recipient, whereas that of a neighbour is fixed to a strategy donor. Some previous studies have considered the different situations \cite{Ohtsuki2006a, Ohtsuki2006c} and showed that the direction of strategy imitation can influence the evolutionary outcomes. For example, one study found that cooperation is enhanced on various lattices if the focal agent is a recipient as opposed to a donor \cite{Wu2009b}. Although strategy updating rules in these studies were all based on the premise that fitter strategies are more likely to proliferate in the population, the ensuing cooperation levels can differ. Hence, because the details of evolutionary processes, such as strategy updating rules, can affect outcomes, the robustness of evolutionary outcomes have been compared between various rules \cite{Zukewich2013}. For example, some studies dealt with models other than the prisoner's dilemma, and investigated whether the direction of copying (imitation) affects fixation probability of an advantageous mutant \cite{Antal2006, Hindersin2015}. In addition, a recent study showed that strategy updating rules can change the consequences of evolutionary processes in well-mixed populations with mutations \cite{Kaiping2014}, and coevolution of strategies and updating rules has been considered \cite{Cardillo2010}. In addition, strategy updating rules that are not informed by imitation of fitter individuals have been investigated \cite{Roca2009}. 

In contrast to the preceding literature that examined the evolutionary process on static networks, our current work investigates the effect of strategy updating rules in coevolutionary games. Although many previous studies assume a specific strategy updating rule, such as VMLD, and showed that cooperation is enhanced in combination with network evolution, the roles of updating rules in these phenomena have not been elucidated. Because network coevolution supports cooperation \cite{Fehl2011}, further studies are warranted to investigate dependences on the details of these models. 

Herein, we study the effect of the combination of link adaptation and three strategy updating rules. Three updating rules used in this study arose from research on network interactions. Node-based strategy updating occurrs under the first two rules: voter-model-like dynamics (VMLD) and invasion-process-like dynamics (IPLD) \cite{Wu2009b}. Specifically, under these two dynamics, one \textit{node} (agent $i$) is selected randomly, and then one neighbour of that node (agent $j$) is selected randomly. Strategy updating occurs by comparing the payoff of these two agents. Under VMLD, agent $i$ copies the strategy of neighbour $j$ with higher probability if agent $j$ earns a larger payoff when compared with agent $i$. In contrast, under IPLD, neighbour $j$ may imitate the strategy of agent $i$. Therefore, a randomly chosen neighbour ($j$) serves as a strategy donor under VMLD and a strategy recipient under IPLD. The last rule is edge-based dynamics (EBD). Under this rule, one \textit{link} ($E_{ij}$) is selected randomly, and the role of the two connected agents (donor or recipient) is randomly assigned. The payoff of these two agents is compared, and a recipient copies the donor's strategy with higher probability if a donor earns a larger payoff. Unlike other two rules, EBD does not fix the roles of the focal agent and its neighbour in strategy transmission, and is eclectic. In the present analyses, VMLD and IPLD are sometimes biased toward enhancing and suppressing cooperation, respectively. Hence, in these cases, EBD with an intermediate feature may serve as a less biased rule. 

Here, we first detail our coevolutionary model and the three strategy update rules. We next report the results of a computer simulation. Lastly, we discuss the implication of our results for the modelling of the evolutionary process of human cooperation.

\section{Model}
Let us assume that $N$ agents are located on a (social) network defined by the neighbours of each node. Links between nodes represent the social relationship. Initially, each agent has the same number of neighbours ($\average{k}$) that are randomly linked to other nodes (see \cite{Santos2005a} for the generation process of this random regular network). Half of the agents, who are chosen randomly, are cooperators, and the rest are defectors. We denote agents' strategy by the two-dimensional vector $s$. Agent $i$ is a cooperator if $s_i = (1, 0)^{T}$ and a defector if $s_i = (0, 1)^{T}$. In this numerical simulation, the payoff matrix ($A$) is given by \cite{Nowak1992}:
\begin{eqnarray}
\bordermatrix{
& C & D \cr
C & 1 & 0 \cr
D & b & 0 \cr
},
\end{eqnarray}
where $b$ is the temptation to defect ($1 < b < 2$). In each time step, a strategy updating event or a partner switching (link adaptation) event occurs. 

Strategy updating events occur with probability $1-w$. We use three types of strategy updating mechanisms. The first one is VMLD. Under this rule, one \textit{node} ($i$) is chosen randomly, and one of the $i$'s neighbours ($j$) is also selected randomly. Then, each agent plays the prisoner's dilemma game with their neighbours and collects a payoff:
\begin{equation}
\Pi_i = \Sigma_{l \in \mathcal{N}_i}{s_i^T A s_l},
\end{equation}
where $\mathcal{N}_i$ is the set of agent $i$'s neighbours and $\Pi_j$ is accumulated in the same manner. The payoff is reset to zero at the end of each event. Agent $i$ decides whether to copy $j$'s strategy based on their accumulated payoff. Specifically, agent $i$ copies the strategy of agent $j$ with a probability calculated by Fermi's rule {\cite{Szabo1998}}:
\begin{equation}
P(s_i \leftarrow s_j) = [1 + \exp(-\beta(\Pi_j - \Pi_i))]^{-1}.
\end{equation}
The value of $\beta$ represents the intensity of selection ($\beta \to 0$ implies random drift, whereas $\beta \to \infty$ implies imitation dynamics). 

The second rule is IPLD. Under this rule, two agents ($i$ and $j$) are selected and accumulate a payoff ($\Pi_i$ and $\Pi_j$) in the same manner as VMLD. The difference is that the neighbour $j$ copies the strategy of agent $i$ with the following probability:
\begin{equation}
P(s_j \leftarrow s_i) = [1 + \exp(-\beta(\Pi_i - \Pi_j))]^{-1}.
\end{equation}
Therefore, the chosen neighbour ($j$) serves as a potential strategy donor under VMLD and a potential recipient under IPLD.

The third rule is EBD. Under this rule, one \textit{edge} is chosen randomly and each agent's role in strategy updating (donor or recipient) is randomly assigned. Each agent plays the prisoner's dilemma with their neighbours and collects a payoff. Next, a recipient ($i$) copies the strategy of the donor ($j$) with probability $\mathrm{P}(s_i \leftarrow s_j)$. Previous study assumed that agent $j$ copies $i$'s strategy unless agent $i$ imitates $j$'s strategy \cite{Fu2009}. In contrast, agent $j$ is not permitted to update the strategy because only one agent can copy the neighbour's strategy under the conditions of the other rules (VMLD and IPLD). Note that both agent $i$ and $j$ have the possibility to copy the strategy of the other agent \textit{ex ante} under EBD, whereas the role in the strategy updating is fixed under VMLD and IPLD.

With probability $w$, partner switching (link adaptation) events occur. The basic idea of the partner switching mechanism used in our current work stems from Zimmermann et al. \cite{Zimmermann2004}. In their pioneering work on the coevolutionary prisoner's dilemma game, these authors reported that cooperation can evolve if a \textit{defective} agent deters the relationship with another \textit{defective} agent. Here, we also assume that edges between two defectors are unstable and that one of the defectors tries to deter the relationship. Specifically, in link adaptation events, one edge {$E_{ij}$} is chosen randomly. If the chosen edge represents a $D-D$ interaction, one randomly chosen agent $i (j)$ stops the interaction with the current partner $j (i)$ and constructs a new link with a randomly chosen agent. If the selected edge represents other situations ($C-C$ or $C-D$), nothing occurs during that period. We impose the restriction that nodes with one link do not lose edges so that they can participate in prisoner's dilemma game.

Notably, we do not directly replicate the model of Zimmermann et al. For example, we use the Fermi function in strategy updating events even though it was not used in the original work because the effect of the intensity of selection is the quantity of interest in our research. In addition, we use asynchronous updating, whereas the original paper used synchronous updating, and agents who change their strategy from $C$ to $D$ deter the relationship. Because of these differences, our results should not be regarded as a validation or a criticism of the work by Zimmerman et al. Here, our goal is to examine the effects of strategy updating mechanisms.

\section{Results}
To investigate the effects of updating rules in coevolving prisoner's dilemma games, we conducted a numerical simulation. Each simulation run continued for $2 \times 10^7$ periods, and the values of the following $10^6$ periods were recorded to compute the average frequency of cooperators, unless one strategy dominated the population. We conducted 1000 independent simulations for each combination of parameters, and calculated the mean values of the simulation results. These values were always utilised unless otherwise stated.

\begin{figure}[tbp]
\centering
\vspace{-5mm}
\includegraphics[width = 110mm, trim= 0 20 0 0]{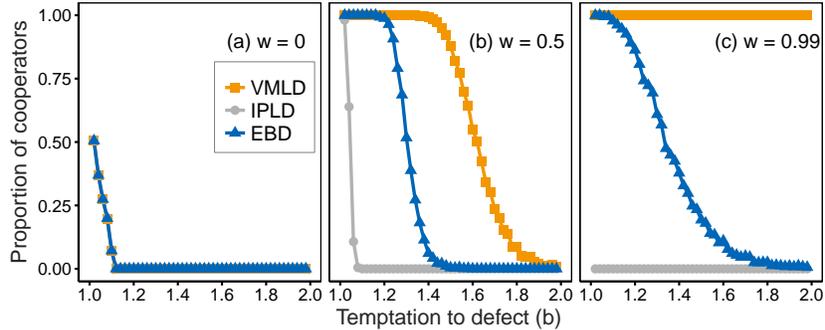}
\caption{\small Proportion of cooperators as a function of the temptation to defect ($b$). In the absence of partner switching, the same results are observed regardless of updating rules ($w = 0$ in panel (a)). In comparison, cooperators with partner switching resist larger temptations to defect in the order of VMLD, EBD and IPLD ($w = 0.5$ in panel (b) and $w = 0.99$ in panel (c)). Parameters: $N = 1000, \average{k} = 8, \beta = 1$.}
\label{fig_pC_b}
\end{figure}

The emerging pattern as a function of the temptation to defect ($b$) is shown in Figure \ref{fig_pC_b}. As expected, a larger temptation to defect induces less cooperation regardless of the strategy updating rule. Here, our interest is to compare the different strategy updating rules. In comparisons of strategy updating rules without partner switching ($w = 0$), cooperation levels do not differ according to strategy updating rules (panel (a)). In contrast, different patterns are observed with the introduction of partner switching ($w = 0.5$). VMLD facilitates the evolution of cooperation under a wider range of temptations to defect, whereas cooperation deteriorates rapidly under IPLD. The results of EBD lie between these two results, and similar patterns are observed with rapid partner switching ($w = 0.99$, panel (c)).

\begin{figure}[tbp]
\centering
\vspace{-5mm}
\includegraphics[width = 135mm, trim= 0 20 0 0]{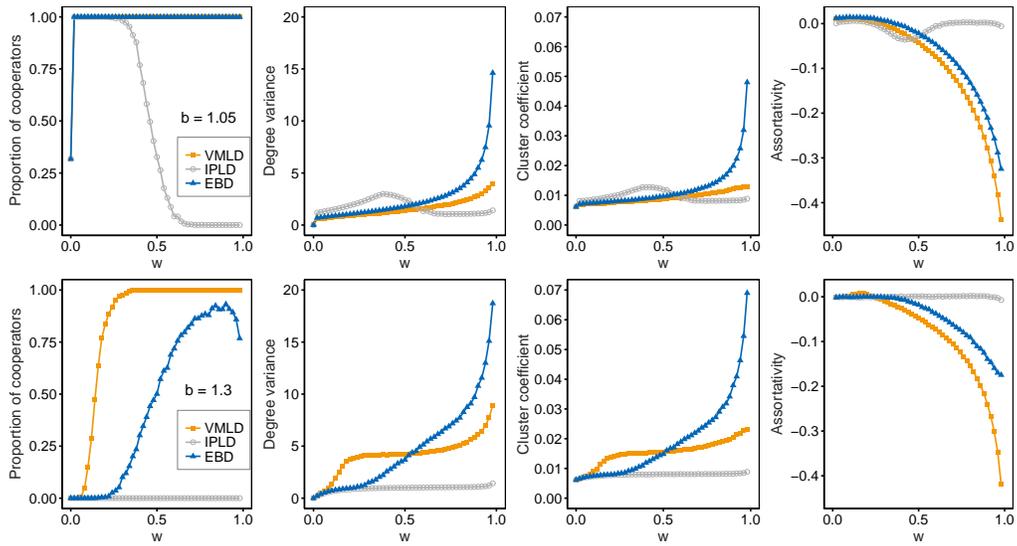}
\caption{\small Proportion of cooperators and resulting network properties as a function of frequencies of link adaptation ($w$). Fast link adaptation facilitates (suppresses) the evolution of cooperation under VMLD (IPLD), whereas partner switching helps cooperators under EBD at appropriate speeds. Larger degree heterogeneity, higher cluster coefficients and negative assortativity tend to coevolve with higher levels of cooperation. Parameters: $b = 1.05$ (upper four panels), $b = 1.3$ (lower four panels); fixed; $N = 1000, \average{k} = 8, \beta = 1$.}
\label{fig_eff_w}
\end{figure}

We next examined the effect of the frequency of partner switching. Figure~\ref{fig_eff_w} shows the resulting proportions of cooperators and network structures. Reported network properties include normalised variance of degrees ([$\average{k_i^2} - \average{k_i}^2]/ \average{k_i}$), cluster coefficients and assortativity \cite{Newman2002}, which were recorded at the end of each simulation run. The upper four panels show results with $b = 1.05$ and the lower panels show results with $b = 1.3$. 

When $b = 1.05$, cooperation is greatly enhanced under VMLD and EBD even with a small probability of link adaptation. In contrast, under IPLD, we observed a non-monotonic relationship between $w$ and the proportion of cooperators. Notably, the evolution of cooperation is \textit{impeded} with a higher frequency of partner switching. When the temptation to defect is larger ($b = 1.3$), the effect of the frequency of link adaptation is consistent with the aforementioned results. Cooperation can evolve easily even with small $w$ under VMLD, whereas partner switching does not help the evolution of cooperation under IPLD. The results for EBD lie between these two results. Under EBD, an extremely larger $w$ deteriorates cooperation, suggesting that defectors can find cooperative partners using larger opportunities for partner switching.

The resultant patterns of networks are compatible with those reported in a previous study of coevolutionary games \cite{Tanimoto2009}. In particular, the first character of the network that coevolves with cooperation is degree heterogeneity. When $b = 1.05$, degree heterogeneity increases with greater opportunities for link adaptation under VMLD and EBD. In contrast, a non-monotonic relationship similar to that of cooperation level is observed under IPLD. When $b = 1.3$, degree heterogeneity increases more with smaller values of $w$ under VMLD than under those of EBD. This pattern corresponds with that of cooperation level, and higher levels of cooperation are achieved with a smaller $w$ under VMLD. Moreover, degree heterogeneity is almost suppressed under IPLD.

Larger cluster coefficients also coevolve with cooperation, especially in VMLD and EBD. Moreover, a non-monotonic pattern is observed again under IPLD when $b = 1.05$. In the study by Zimmermann et al. \cite{Zimmermann2004} modest evolution of cluster coefficients was observed with enhanced cooperation. Other studies also show that larger cluster coefficients are related to the evolution of cooperation \cite{Roca2009}. Furthermore, we find that negative assortativity of degrees emerges with cooperation, and emerging negative assortativity was previously shown in a study of coevolution using the prisoner's dilemma game \cite{Tanimoto2009}.

Figures \ref{fig_pC_b} and \ref{fig_eff_w} show that cooperators flourish in the order of VMLD, EBD and IPLD. We can explain this pattern by considering how often the cooperators are chosen as the potential strategy donor or recipient. In the coevolving prisoner's dilemma, it is often assumed that defectors are more likely to lose edges. Cooperators tend to have larger number of neighbours, and therefore cooperators are more likely to be chosen as the \textit{neighbour} of the focal agent under VMLD and IPLD. Under VMLD, because the randomly chosen neighbour serves as the potential donor of the strategy, cooperators have a greater chance to be imitated. This advantage for cooperators was previously observed, i.e., the result that cooperators flourish despite the smaller average payoff when compared with defectors \cite{Cong2014}. In contrast, because the neighbour serves as the potential strategy recipient under IPLD, cooperators have more opportunity to imitate others' strategy. This difference in the frequency of cooperators becoming a donor (recipient) of the strategy supports (hinders) the evolution of cooperation under VMLD (IPLD). Because cooperators have a greater chance to imitate and to be imitated under EBD, the result lies between the other two processes.

\begin{figure}[tbp]
\centering
\vspace{-5mm}
\includegraphics[width = 100mm, trim= 0 20 0 0]{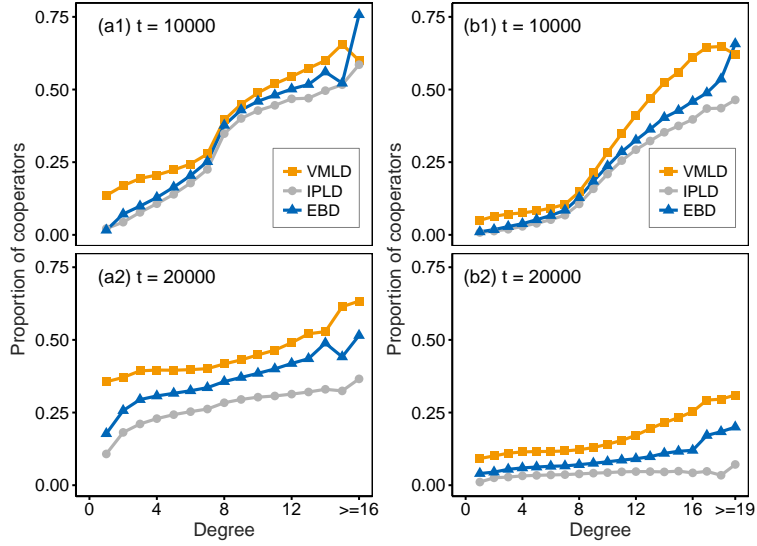}
\caption{\small Frequency of cooperators at $t = 10^4$ and $t = 2 \times 10^4$ as a function of nodes' degree at $t = 10^4$. Accumulations of $10^4$ simulation runs are presented. Because nodes with larger degrees rarely appear during simulations and the sample size is small, we report the results of large degrees together. (a1) and (a2), Larger degree facilitates cooperation regardless of strategy updating rules when $b = 1.05$ and $w = 0.3$. (b1) and (b2), Cooperators can exploit the benefits of larger degrees in the order of VMLD, EBD and IPLD when $b = 1.3$ and $w = 0.5$. Fixed parameters: $N = 1000, \average{k} = 8, \beta = 1$.}
\label{fig_pC_k}
\end{figure}

This tendency influences how the cooperators enjoy the benefit of a larger degree. In Figure \ref{fig_pC_k}, we classified agents by their degrees ($k_i$) at $t = 10^4$ and calculated the proportion of cooperators at $t = 10^4$ and $t = 2 \times 10^4$, respectively. Panels (a1) and (b1) show proportions of cooperators at $t = 10^4$, and panels (a2) and (b2) show values at $t = 2 \times 10^4$. In typical coevolving prisoner's dilemma games, cooperators achieve larger degrees and hence have greater resistance to invasion by defectors. This relationship is confirmed in panels (a1) and (a2), which show results for $b = 1.05$ and $w = 0.3$. Under these conditions, almost full cooperation is observed (upper left panel of Figure \ref{fig_eff_w} shows the corresponding final outcomes) and agents with larger degrees are more likely to be cooperators. Furthermore, these agents remain cooperative at $t = 2 \times 10^4$ regardless of the strategy updating rules, although the relationship between the frequency of cooperators and each agent's degree is weakened. 

When $b = 1.3$ and $w = 0.5$, the relationships vary depending on strategy updating rules (lower left panel of Figure \ref{fig_eff_w} shows the corresponding final cooperation level). In panel (b1), large degrees are related to higher cooperation levels regardless of updating rules, and the relationship is stronger in the order of VMLD, EBD and IPLD. However, the data in panel (b2) indicates that whether an agent with large degree can \textit{remain} cooperative depends on updating rules. Under VMLD, in which full cooperation was achieved, the same pattern is observed and larger degrees facilitate cooperators. In contrast, under IPLD, in which defectors were dominant, relationships are unstable and cooperators do not enjoy the benefits of larger degrees. Under EBD, in which moderate levels of cooperation were observed, the results lie between those of the other two rules. Specifically, the relationship between degree and cooperation is confirmed but is weaker than that under VMLD.

\begin{figure}[tbp]
\centering
\vspace{-5mm}
\includegraphics[width = 60mm, trim= 0 20 0 0]{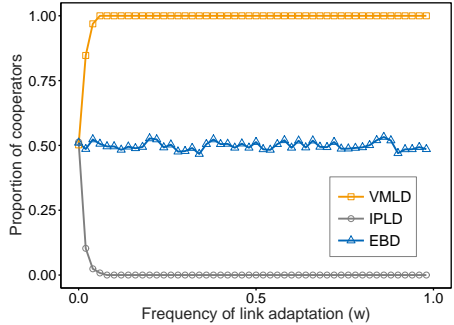}
\caption{\small Fractions of cooperators when the payoff has neutral effect ($\beta = 0$). Partner switching helps (hinders) cooperation under VMLD (IPLD). Neither cooperation nor defection is favored under EBD. Parameters: $N = 1000, \average{k} = 8, b = 1.3, \beta = 0$.}
\label{fig_pC_w}
\end{figure}

Different patterns also appear for the effect of the intensity of selection ($\beta$). When $\beta = 0$, the payoff from games has a totally neutral effect on evolution. As is displayed in Figure \ref{fig_pC_w}, full cooperation is achieved under VMLD with a small opportunity for partner switching, whereas cooperation is suppressed under IPLD. The results under EBD lie between these results, i.e., neither cooperation nor defection is favoured. Because cooperators acquire larger degrees, they have larger opportunities to become strategy donors or recipients under VMLD or IPLD, respectively. When the payoff has no effect on evolutionary outcomes, this difference in directions of strategy imitation is directly reflected by cooperation levels. 

Thus, VMLD, which has been commonly used in previous studies \cite{Wu2009b}, can help the evolution of cooperation independently of the payoff because cooperators are more likely to be chosen as the strategy donor. Some studies showed that weak selection favours cooperation in coevolving games \cite{Santos2006, Fu2007, Yang2013a}. We infer that VMLD also played an important role in the evolutionary process, in addition to a coevolutionary mechanism. Hence, VMLD and IPLD are biased toward supporting and suppressing cooperation, respectively, whereas EBD has a neutral effect on cooperation levels in coevolving prisoner's dilemma games. 

\begin{figure*}[tbp]
\centering
\vspace{-5mm}
\includegraphics[width = 130mm, trim= 0 20 0 0]{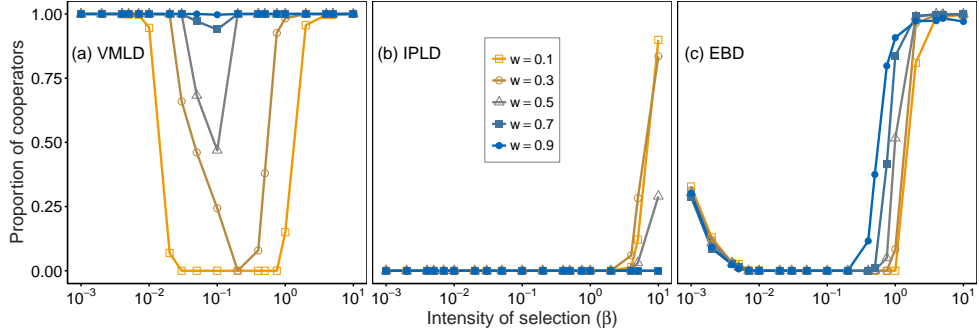}
\caption{\small Fractions of cooperators as a function of the intensity of selection ($\beta$). Non-monotonic relationship is observed and moderate intensity of selection hinders the evolution of cooperation under VMLD and EBD. Only strong selection helps cooperators under IPLD. Parameters: $N = 1000, \average{k} = 8, b = 1.3$.}
\label{fig_pC_beta_w}
\end{figure*}

We also show the effect of the intensity of selection ($\beta$) in Figure \ref{fig_pC_beta_w}. Under VMLD, full cooperation is achieved more easily when selection is weak or strong, and a higher frequency of link adaptation is required with moderate values of $\beta$. Qualitatively the same pattern is observed under EBD; cooperation deteriorates when the values of $\beta$ are moderate, and a higher $w$ enhances the evolution of cooperation even with a weaker intensity of selection. This pattern is contrary to the results under the \textit{static} network, which showed that there exists an optimal intensity of selection in supporting evolution of cooperation \cite{Szabo2005}. Under IPLD, stronger selection is required for cooperators to survive, especially when the high frequency of link adaptation hinders the evolution of cooperation. 

The overall resulting pattern can be explained by the combination of the effects of partner switching and strategy updating rules. Partner switching is beneficial for cooperators due to the emerging degree heterogeneity \cite{Szolnoki2008b} and cooperators can gain a larger accumulated payoff by link adaptation unless $w$ is too large (see Figure \ref{fig_eff_w}). As a result, a stronger intensity of selection helps cooperators with a larger degree to maintain their strategy. However, because cooperators are sustained by VMLD (and to a lesser extent by EBD) when the intensity of selection is extremely weak, a stronger intensity of selection can reduce the advantage of the strategy updating rules. Indeed, a non-monotonic effect of $\beta$ is observed under VMLD and EBD. In contrast, IPLD hinders the evolution of cooperation, therefore only the combination of a stronger intensity of selection and appropriate frequency of partner switching helps cooperators.

Because the results of evolutionary network games can vary with the synchronicity of strategy updates, we perform further computations with synchronous updating. In this version of the model, each simulation round is conducted by selecting all $N$ agents in a random order. Under these conditions, link adaptation events occur with probability $W$, and a focal agent may cut the link with a randomly selected neighbour and reconnect it with a randomly selected agent. Link adaptation occurs when a focal agent and a selected neighbour are both defectors. Conversely, strategy updates occur with probabilities of $1-W$ and lead to accumulations of payoffs for a focal agent and a single neighbour as in asynchronous updating. Strategy imitations are determined using Fermi's rule. In VMLD, a focal agent becomes a recipient of the strategy and a neighbour becomes a donor of the strategy, and the roles of the two agents are switched in IPLD. Under random (RAND) rule, roles of two agents are determined randomly. We introduce RAND rule instead of EBD because sweeping is conducted by agent after agent. This rule corresponds to EBD where the roles of agents are not fixed.

Some clarifications may be required before explaining the simulation results. A single link can be severed by both agents who are connected by that edge at one time step, whereas only one decision is reflected in the next round. In addition, although some agents may serve as recipients of the strategy multiple times under IPLD and RAND, the consequence of one event is reflected in the next round, regardless of whether imitation occurred in that event.

\begin{figure}[tbp]
\centering
\vspace{-5mm}
\includegraphics[width = 110mm, trim= 0 20 0 0]{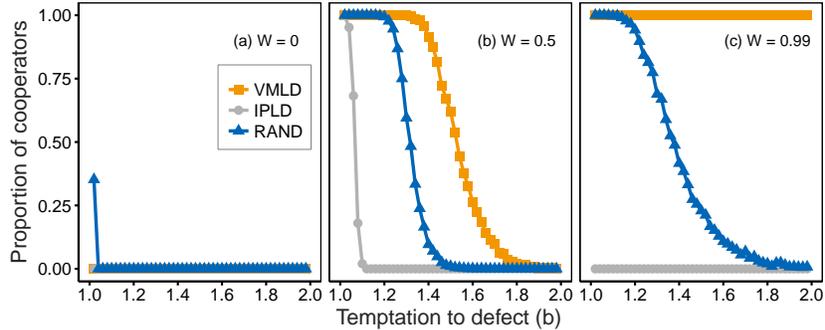}
\caption{\small Proportion of cooperators as a function of the temptation to defect ($b$). 
Simulation runs continued for $2 \times 10^4$ periods, and the values of the following $10^3$ periods were recorded. Due to slow convergence, sampling started after $4 \times 10^5$ periods when $W = 0$ and $b \leq 1.04$. The pattern in Figure \ref{fig_pC_b} is replicated with synchronous updating. Parameters: $N = 1000, \average{k} = 8, \beta = 1$.}
\label{fig_pC_b_sync}
\end{figure}

These simulations with synchronous updating (Figure \ref{fig_pC_b_sync}) show differing behaviours without link adaptation (panel (a)) and cooperators are more likely to survive under the RAND rule. However, a similar pattern to that observed in Figure \ref{fig_pC_b} is replicated when partner switching is permitted and cooperation is supported in the order of VMLD, RAND and IPLD (panels (b) and (c)). This is because the same logic also works in synchronous updating. Under VMLD, cooperative agents who have large degrees are more likely to be selected as a neighbour of a focal agent and have better opportunities to enforce their strategy. In contrast, cooperative agents are more likely to imitate neighbour's strategies under IPLD and defectors who tend to have small numbers of neighbours may never serve as strategy recipients in single time steps. The results of the RAND rule which has the characteristics of VMLD and IPLD lie between the other two cases.

Finally, we examine the effects of strategy updating rules on games conducted using static heterogeneous networks. Because no link adaptation occurs under these conditions, we cannot directly apply the same logic that affected the evolutionary processes of coevolutionary games. By the definition of \textit{static} networks, the adopted strategy does not change degrees, and the probabilities of agents becoming recipients or donors of the strategy remain unaffected. However, a seminal study of the positive effects of degree heterogeneity \cite{Santos2006} suggested that hub nodes are more likely to be cooperators and that clusters of cooperators tend to form around them. These hub nodes may contribute to the evolution of cooperation in combination with VMLD. Because hub agents are more likely to be donors of strategies under VMLD, cooperation may be efficiently enforced. In this regard, studies of the effect of updating rules on the games on heterogeneous networks may be a natural extension of this study. The present heterogeneous network is generated using preferential or uniform attachment, corresponding with evolutionary processes on Barab\'asi-Albert (BA) networks \cite{Barabasi1999} and exponential networks, respectively.

\begin{figure}[tbp]
\centering
\vspace{-5mm}
\includegraphics[width = 95mm, trim= 0 20 0 0]{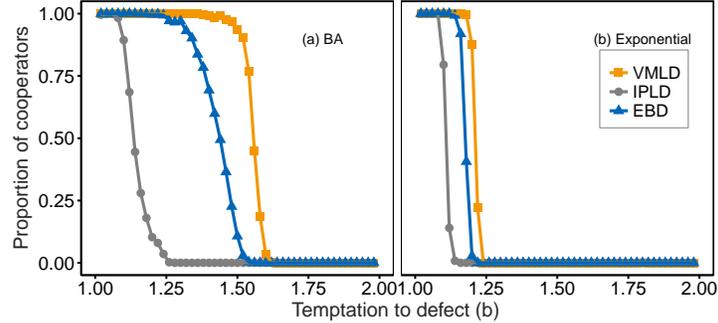}
\caption{\small Proportion of cooperators as a function of the temptation to defect ($b$) on static heterogeneous networks. Initially, we generated a complete network comprising $m_0 = 6$ nodes and $(N - m_0)$ nodes were connected with $m = 6$ nodes by preferential (panel(a)) or uniform (panel (b)) attachments. Simulation runs continued for $2 \times 10^8$ ($2 \times 10^7$) periods, and the values of the following $10^7$ ($10^6$) periods were recorded with BA (exponential) networks. Cooperators are more likely to proliferate in the order of VMLD, EBD and IPLD, and this tendency is stronger with BA networks. Parameters: $N = 1000, w = 0, \beta = 1$.}
\label{fig_static}
\end{figure}

The results of the present simulation of heterogeneous networks (Figure~\ref{fig_static}) show that VMLD offer the most advantageous environment for cooperators. In contrast, cooperators can flourish in IPLD under limited parameter ranges. Because agents with large degrees are more likely to become recipients of strategies under IPLD, it becomes more difficult for hub nodes to remain cooperative and establish clusters of cooperators. Cooperation levels under EBD lie between those of the other models. In addition, comparisons with BA (panel (a)) and exponential (panel (b)) networks show that differences that emerge from varied strategy updating rules are larger among BA networks. BA networks have larger degree heterogeneity, leading to increased frequencies of becoming a donor or recipient. Compared with coevolving games, different patterns emerge under assumptions of extremely weak selection. Specifically, significant differences between updating rules are not observed when $\beta = 0$ (see Figure 4 for the results of coevolving games) and the cooperation level is about 0.5 under all three rules, despite the presence of degree heterogeneity (data not shown). Constant degrees of each node precluded effects of strategies on frequencies of becoming a donor or a recipient, whereas in coevolving games, cooperativeness increases the opportunity for strategy enforcement or learning in combination with VMLD or IPLD, respectively. In static networks, VMLD supports the formation of cooperative clusters that tend to produce large payoffs, and this leads to the evolution of cooperation.

\section{Discussion}
In this paper, we compare the three strategy updating rules, VMLD, IPLD and EBD, on the coevolutionary prisoner's dilemma game. Our results show that VMLD, which were adopted in many previous studies, favour cooperation under a wide range of parameters. This is because cooperators who have larger number of neighbours are more likely to become a potential strategy donor. In contrast, cooperators are more likely to become a strategy recipient under IPLD, which prevents cooperators from enjoying the benefit of degree heterogeneity. Consequently, a higher frequency of partner switching sometimes deters the evolution of cooperation under IPLD. The results of EBD lie between these two outcomes. Larger degree heterogeneity, higher clustering, and negative assortativity coevolve with cooperation. In addition, cooperation is supported or suppressed irrespective of payoffs under VMLD or IPLD, respectively. Hence, combinations of strategy updating rules and network evolution can affect evolutionary outcomes independently of game results. EBD with intermediate characters are less amenable to this effect. Furthermore, combined with the effect of the strategy updating rules, a non-monotonic relationship between the intensity of selection and proportion of the cooperators is observed under VMLD and EBD, whereas stronger selection favours cooperation under IPLD. Similar patterns of enhanced cooperation are observed with synchronous updating and heterogeneous static networks.

In previous studies, VMLD was often utilised as the strategy updating rule. In the context of modelling human behaviour, strategy updating rules can be regarded as the assumption for the social learning (imitation) process. Specifically, updating rules determine the direction of influence. In our simulation, strategy updating rules influenced the possibility that agents with a larger degree become a strategy donor (or recipient) and thus affected the resulting cooperation level. Therefore, understanding who will be more likely to imitate others and who will be imitated by others may be very important for studying the evolution of human cooperation.






\end{document}